\documentclass[preprint,nofootinbib,aps,10pt,twocolumn,superscriptaddress]{revtex4} 
\usepackage[dvips]{graphicx}

\usepackage{array,amsmath,amsthm,feynmf}
\usepackage{bm}
\usepackage{slashed}
\usepackage{times}
\usepackage{epsfig}
\usepackage{graphicx}
\usepackage{color}
\usepackage{cancel}
\usepackage{amssymb}
\usepackage{textcomp}

\definecolor{dGreen}{rgb}{0,.6,0}
\newcommand{\be}{\begin{equation}}
\newcommand{\ee}{\end{equation}}
\newcommand{\bea}{\begin{eqnarray}}
\newcommand{\eea}{\end{eqnarray}}

\newcommand{\lsim}{\buildrel < \over {_\sim}}

\begin{document}

\smash{\hspace{5.5 cm}NPAC-13-02}
\smash{\hspace{5.5 cm}CALT-68-2919}\vspace{-10cm}


\title{\Large Color Breaking in the Early Universe}

\vspace{4.0cm}
\author{Hiren H. Patel}
\email{hhpatel@wisc.edu}
\affiliation{
{University of Wisconsin-Madison, Department of Physics} \\
{1150 University Avenue, Madison, WI 53706, USA}}
\author{Michael J. Ramsey-Musolf}
\email{mjrm@physics.wisc.edu}
\affiliation{
{University of Wisconsin-Madison, Department of Physics} \\
{1150 University Avenue, Madison, WI 53706, USA}}
\affiliation{
{California Institute of Technology}\\
{Pasadena, CA 91125 USA}}
\author{Mark B. Wise}
\email{wise@theory.caltech.edu}
\affiliation{
{California Institute of Technology}\\
{Pasadena, CA 91125 USA}}
\date{\today}
\begin{abstract}
We explore the possibility that SU(3)$_C$ was not an exact symmetry at all times in the early universe, using minimal  extensions of the standard model that contain a color triplet scalar field and perhaps other fields. We show that, for a range of temperatures,  there can exist a phase in which the free energy is minimized when the color triplet scalar has a non-vanishing vacuum expectation value, spontaneously breaking color. At very high temperatures and at lower temperatures color symmetry is restored. The breaking of color in this phase is accompanied by the spontaneous breaking of $B-L$ if the color triplet scalar Yukawa couples to quarks and/or leptons. We discuss the requirements on the minimal extensions needed for consistency of this scenario with present collider bounds on new colored scalar particles.
\end{abstract}
\maketitle
\section{Introduction}
The discovery of a boson with mass around $125$ GeV  at the Large Hadron Collider (LHC) whose properties are so far roughly consistent with those expected for the Standard Model (SM) Higgs boson\cite{Aad:2012tfa,Chatrchyan:2012ufa} adds new credibility to the paradigm of scalar field-driven spontaneous symmetry breaking in  the early universe. In the Standard Model Higgs mechanism paradigm, the electroweak $\text{SU(2)}_\text{L}\times$U(1)$_Y$ symmetry breaks to electromagnetic U(1)$_\mathrm{EM}$ symmetry in a single step at a temperature, $T\sim\mathcal{O}(100)$ GeV. Monte Carlo studies indicate that if only SM fields are present, the electroweak phase transition (EWPT) at these temperatures would be  cross over  for a Higgs boson with mass indicated by the LHC observation \cite{Gurtler:1997hr,Laine:1998jb,Csikor:1998eu,Aoki:1999fi}. In this case, one would expect little else in the way of other cosmological implications or related experimental signatures, such as the production of relic gravity waves were the EWPT to have been first order.

It is interesting, then, to ask whether electroweak symmetry breaking (EWSB) needs to have followed the simple trajectory implied by the SM. For many years, theorists have explored the possibility that the presence of additional scalar fields might change the character of the EWPT, making it strongly first order as required for electroweak baryogenesis or the generation of potentially observable relic gravity waves (for a recent review and references, see Ref.~~\cite{Morrissey:2012db}). In most of these studies, electroweak symmetry breaking (EWSB) still proceeds through a single step, though a few studies of multistep transitions have appeared in the literature\cite{Land:1992sm,Hammerschmitt:1994fn,Profumo:2007wc,Patel:2012pi}. In all cases, the focus has been on EWSB while avoiding the loss of $\text{SU(3)}_\text{C}$ that is known to be a good symmetry today. In the minimal supersymmetric Standard Model (MSSM), for example, the avoidance of color-breaking (CoB) and charge-breaking (ChB) vacua imposes severe constraints on the parameters of the scalar potential when one invokes the scalar superpartners of the right-handed top quarks to obtain a strong first order EWPT. In this case, the universe transitions to a color-conserving EW vacuum is that metastable, and for a sufficiently light Higgs boson, remains there as the rate for tunneling to a deeper  CoB/ChB vacuum is adequately suppressed\cite{Carena:2008vj}. It has been proposed in Ref.~\cite{Stojkovic:2007dw} that the  energy of a color-conserving EW false vacuum may be responsible for the cosmic acceleration.

It is possible, however, that the early universe underwent an epoch when $\text{SU(3)}_\text{C}$ was spontaneously broken but later restored. To our knowledge, there exist no theoretical arguments or experimental observations that would preclude this possibility. If realized, it would mean that today's symmetries need not have been present at all times in the early universe, contrary to the conventional picture in which symmetries are only lost and not restored as the universe cools. 

We are not the first to consider the possibility that part of the the zero temperature unbroken gauge group of the standard model or other gauge theory might have been broken in the early universe.  Weinberg first observed that in the context of an O(n)$\times$O(n) gauge theory, one may encounter a transition to a state of lower symmetry O(n)$\times$O(n)$\to$ O(n)$\times$O(n-1) with increasing $T$ \cite{Weinberg:1974hy}. Langacker and Pi subsequently applied this idea to show how breaking of  U(1)$_\mathrm{EM}$ and its subsequent restoration could provide a solution to the monopole problem in grand unified theories (GUTs) \cite{Langacker:1980kd}. The authors of Refs.~\cite{Dvali:1995cj,Dvali:1996zr} proposed an alternate solution for an SU(5) GUT that relied on non-restoration  of the symmetry at high-$T$ in the presence of a suitable scalar representation of the gauge group. Similar ideas were explored for spontaneous breaking of CP-invariance in Refs.~\cite{Mohapatra:1979qt,Mohapatra:1979vr}. For the specific case of SU(3)$_C$, the authors of Ref.~\cite{Cline:1999wi} studied the possibility that a CoB phase preceded the color-conserving EW phase in the MSSM and found that without the introduction of additional interactions, a universe that cooled into a CoB phase would always remain there. 

In this note, we demonstrate
how $\text{SU(3)}_\text{C}$-breaking followed by its restoration might arise, using minimal extensions of the SM scalar sector. We further show that this scenario may be realized without the occurrence of a metastable CoB vacuum at $T=0$.
We also outline some of the phenomenological consequences and constraints, focusing on two representative cases corresponding to the SM plus: 
\begin{itemize}
\item[(a)] a single multiplet of sub-TeV colored scalars $C$ 
\item[(b)] a single multiplet of colored scalars $C$, as well as a gauge singlet $S$
\end{itemize}

In both cases, the basic mechanism for achieving the novel pattern of symmetry breaking where color is broken in an intermediate phase involves a negative mass-squared term for the colored scalar in the scalar potential. A positive contribution to its $T=0$ physical mass squared, $m_C^2$,  is generated by quartic terms of the form $C^\dag C \varphi^\dag\varphi$, where $\varphi$ denotes one of the other scalars in the theory such as the Higgs doublet $H$ or $S$.  Provided the product of the associated quartic coupling and square of the vacuum expectation value (vev) of $\varphi$ is sufficiently large at $T=0$,  the positive contribution to $m_C^2$ overwhelms the negative bare mass squared term, leaving at least a local minimum where color is unbroken at $T=0$.  If the vev of $\varphi$  falls sufficiently rapidly as $T$ increases, an intermediate phase where the minimum of the free energy breaks $\text{SU(3)}_\text{C}$ can exist.

For case (a) $\varphi$ is the Higgs doublet and we find that in order to  maintain perturbative couplings, the colored scalar $C$  cannot be too heavy.  Constraints from LHC new particle searches then imply that its couplings to first and second generation SM fermions must be suppressed. Case (b), wherein $\varphi$ is the singlet,  allows one to avoid this requirement by pushing the mass of $C$ above present LHC bounds while maintaining perturbativity. This is possible because the vev of $S$ can be larger than the Higgs vev, allowing it to give the dominant positive contribution to the $m_C^2$.  As we discuss below, for this case the region of parameter space that achieves a CoB phase for a range of temperatures is restricted not so much by present phenomenological constraints but by the behavior of the scalar potential as a function of $T$. Nevertheless, both cases illustrate the more general possibility of the novel symmetry-breaking pattern of interest here and the considerations that determine its viability\footnote{In analyzing these cases, we have not imposed any requirements on the character of the phase transitions. Instead, we simply concentrate on the possible existence of a CoB phase that is subsequently restored. 
}. At the end of this note, we comment on other variants that may lead to a similar symmetry-breaking pattern.

Before proceeding, we emphasize that the exploratory nature of our study. As with any analysis of the phase history of the universe in perturbation theory, one should take the precise numerical results with an appropriate grain of salt. In the case of the non-restoration of SU(5) gauge symmetry note above, for example, the work of Ref.~\cite{Bimonte:1995xs} indicated that inclusion of higher-order contributions to the effective potential could modify conclusions based on one-loop result\footnote{For a study of the impact of two-loop effects with colored scalars on character of an EWPT, see Ref.~\cite{Cohen:2011ap}}. On the other hand, comparison of lattice results in the standard model and MSSM with perturbative computations suggest that the latter can faithfully reproduce the parametric dependence of phase transition dynamics even if perturbative results for the critical temperatures, latent heat, {\em etc.} are not numerically accurate (for a detailed discussion, see Ref.~\cite{Patel:2011th}). Thus, we suspect the viability of a period of color-breaking and restoration in the minimal models discussed here  -- as well as the general behavior of this possibility as a function of the various interactions --  should hold after inclusion of higher-order contributions or completion of non-perturbative computations. 

Our discussion of this paradigm is organized as follows. In Section \ref{sec:twofield} case (a) is analyzed in detail, while in Section \ref{sec:singlet} we treat case (b). We summarize in Section
\ref{sec:conclude}.  

\section{Two field scenario}
\label{sec:twofield}

To set the notation, we first consider case (a) with $C$ being an $\text{SU(3)}_\text{C}$ triplet. In order to avoid the possibility of   stable color triplet scalar degrees of freedom the field  $C$  must be charged  under $\text{SU(2)}_\text{L}\times\text{U(1)}_Y$ to permit couplings to SM fermions. For example the $C$  could be an $\text{SU(2)}_\text{L}$ doublet  with hypercharge  $Y= 7/6$. Then it has the following Yukawa type interactions with the quarks,
\be
L_Y={C} {\bar u_R} g_{uL} L_L  +  {C}{\bar Q_L} g_{Qe}e_R+{\rm h.c.}\,.
\ee

Here we have suppressed the color, weak isospin and flavor indices. The Yukawa coupling  matrices $g_{ue}$ and $g_{Qe}$ could be such that the dominant decay mode of the colored scalar $ C$ is to a top quark and an anti tau lepton, thereby alleviating present LHC constraints on new colored scalars. Assigning a baron number $1/3$  to $C$ the  tree level lagrangian conserves baryon number\footnote{Non perturbative quantum effects violate this symmetry because it is anomalous.}.  

To make the analysis of the phase structure of the theory as simple as possible we consider instead the illustrative situation where $C$ is an electroweak singlet but color triplet. Then it has no renormalizable interactions with quarks and leptons, and the most general gauge invariant renormalizable potential is\footnote{If the $C$ were also a weak doublet, other terms like $H^\dag\tau^a H C^\dag\tau^a C$ would appear in the scalar potential.}
\begin{align}
\label{eq:modela}
\nonumber V &=  -\mu_H^2 (H^\dag H) -\mu_C^2 (C^\dag C) 
+\frac{\lambda_H}{2} (H^\dag H)^2\\
&\qquad + \frac{\lambda_C}{2} (C^\dag C)^2  
+\lambda_{HC} (H^\dag H) (C^\dag C)\,.
\end{align}
We take all the parameters $\mu_H^2$, $\mu_C^2$, and $\lambda_{H,\,C,\,HC}$ positive. 

The $T=0$ vacuum structure follows from the minimization conditions: 
\be
\label{eq:mincond}
\partial V/\partial\phi^\dag = 0 \qquad \mathrm{for}\qquad \phi = H,\, C\ \ \ .
\ee
 Four possible stationary points emerge, with the corresponding vacuum energies labeled $E_a$, with $a=\{0,\,H,\,C,\,HC\}$:
\begin{itemize}
\item[{\bf (1)}] $H=C = 0$  with energy  $ E_0 = 0$. This is always a local maximum since very small fluctuations in the fields about this point contribute negatively to the potential energy.
\item[{\bf (2)}] Break $\text{SU(2)}_\text{L}$ but not $\text{SU(3)}_\text{C}$:  $\langle H\rangle =(0,\,v_H)^\text{T}$ and $\langle C \rangle =0$. Solving Eq.~(\ref{eq:mincond}) gives
\begin{equation}
\label{eq:vH}
v_H^2={\mu_H^2 \over \lambda_{H}}
\end{equation}
with vacuum energy density 
 \begin{equation}
 \label{eq:EH}
 E_H= -{\mu_H^4 \over 2 \lambda_{H}}
 \end{equation}
 \item[{\bf (3)}] Break $\text{SU(3)}_\text{C}$ but not $\text{SU(2)}_\text{L}\times\text{U(1)}_Y$: $\langle  H \rangle =0$ and  $\langle C \rangle =(0,\,0,\,v_C)^\text{T}$, where, without loss of generality, we have taken the third component of $C$ to acquire a vev.  Then $\text{SU(3)}_\text{C}$ is spontaneously broken to an $\text{SU(2)}$ color subgroup so that in the broken phase there are still three massless gluons but five that get mass by the Higgs mechanism. The solution of Eq.~(\ref{eq:mincond}) then implies
\begin{equation}
\label{eq:vC}
v_C^2={\mu_C^2 \over \lambda_{C}}
\end{equation}
 with energy at the stationary point
\begin{equation}
\label{eq:EC}
 E_C= -{\mu_C^4 \over 2 \lambda_{C}}\ \ \ .
\end{equation}
\item[{\bf (4)}] Break  both $\text{SU(3)}_\text{C}$ and  $\text{SU(2)}_\text{L}\times\text{U(1)}_Y$:
 $\langle H\rangle =(0,\,v_H)^\text{T}$, and $\langle C \rangle =(0,\,0,\,v_C)^\text{T}$,
 with
 \begin{equation}
 \label{eq:vHC}
 v_H^2={\lambda_C \mu_H^2-\lambda_{HC} \mu_C^2 \over \lambda_H \lambda_C-\lambda_{HC}^2},~~~v_C^2={\lambda_H \mu_C^2-\lambda_{HC} \mu_H^2 \over \lambda_H \lambda_C-\lambda_{HC}^2}\ \ \ ,
\end{equation}
leading to an energy density
\begin{equation}
\label{eq:EHC}
E_{HC}=-{1 \over 2}{\lambda_C \mu_H^4+\lambda_H \mu_C^4 -2 \lambda_{HC}\mu_H^2 \mu_C^2 \over \lambda_H \lambda_C - \lambda_{HC}^2}\ \ \ .
\end{equation}
\end{itemize}

The absolute minimum must correspond to (2), implying first  that the eigenvalues $m_H^2$ and $m_C^2$ of the mass-squared matrix, which follows from second derivatives of the potential, must be positive at $C=0$ and $v_H=\mu_H/\sqrt{\lambda_H}$:
\bea
\label{eq:pos1}
m_h^2 = 2 \mu_H^2 = 2\lambda_H v_H^2 & > & 0\\
\label{eq:pos2}
m_C^2 = -\mu_C^2 +\lambda_{HC} v_H^2 & > & 0
\eea
The requirement in Eq.~(\ref{eq:pos1}) is trivially satisfied while Eq.~(\ref{eq:pos2}) constrains the parameters in the potential to satisfy,
\be
\label{eq:pos3}
\lambda_{HC} v_H^2 > \mu_C^2 \,. 
\ee
Eq. (\ref{eq:pos2})  places the upper bound on $m_C$ as a function of $\lambda_{HC}$, 
\be
\label{bound}
m_C < \big({\sqrt{\lambda_{HC}}}\big) v_H \simeq  (174\text{ GeV}) {\sqrt{\lambda_{HC}}}\,.
\ee
If the color triplet complex scalar has generic decay modes to light quarks and leptons, LHC data implies that its mass must be greater than about $500~{\rm GeV}$. Eq.~(\ref{bound})  then implies that $\lambda_{HC}$ is greater than about $9$, taking us into the strong coupling regime.

In addition, the extrema (3) or (4) may also be minima, with positivity conditions. For purposes of illustration, we consider the positivity conditions for (3), denoting the corresponding scalar masses at this point as ${\tilde m}_H$ and ${\tilde m}_C$:
\begin{align}
\label{eq:pos4}
{\tilde m}_H^2 & =  - \mu_H^2+\lambda_{HC}\, v_C^2= -\lambda_H v_H^2 + \lambda_{HC} v_C^2  >  0\\
\label{eq:pos5}
{\tilde m}_C^2 & =  2\mu_C^2=2\lambda_C v_C^2  >  0\,.
\end{align}
Again, the (\ref{eq:pos5}) is trivially satisfied, while (\ref{eq:pos4}) implies that
\be
m_H^2<2 \lambda_{HC} v_C^2  = \frac{2\lambda_{HC} \mu_C^2}{\lambda_C} =  \frac{2\lambda_{HC}(\lambda_{HC}v_H^2-m_C^2)}{\lambda_C}\,.
\ee

\subsection{Symmetry breaking at finite $T$}
\label{sec:modelaT}
We now determine the conditions under which the extremum (2) is the absolute minimum at $T=0$ but either (3) or (4) becomes deeper as $T$ is increased. To that end, we first determine the temperatures $T^\ast$ above which extrema disappear along the $H$ and $C$ directions. For purposes of both simplicity and intuition, our analytic work is based on  the high-$T$ effective theory, wherein the effects of temperature are to replace the coefficients of the quadratic terms in the free energy by their Debye masses:
\bea
\mu_H^2(T)&=&\mu_H^2-T^2\left({ \lambda_H \over 4} +{\lambda_{HC}\over 4}+{3g_2^2 \over 16} +{y_t^2 \over 4}\right) \\
\mu_C^2(T)&=&\mu_C^2-T^2\left({ \lambda_C \over 3} +{\lambda_{HC}\over 6}+{g_3^2 \over 3} \right)\, ,
\eea
where we have neglected the small hypercharge contribution to $\mu_H^2(T)$ and where we recall that the top quark mass is given by
$m_t=y_t v_H$ in our normalization of the Higgs vev. Symmetry will be restored along the $H$ and $C$ directions for $T> T^\ast_{H}$ and $T^\ast_C$, respectively:
\bea
\left(T_H^\ast\right)^2 & = &\frac{ \lambda_H v_H^2}{\left({ \lambda_H \over 4} +{\lambda_{HC}\over 4}+{3g_2^2 \over 16} +{y_t^2 \over 4}\right)}\\
\left(T_C^\ast\right)^2  & = & \frac{\mu_C^2}{\left({ \lambda_C \over 3} +{\lambda_{HC}\over 6}+{g_3^2 \over 3} \right)}\,.
\eea

To obtain the pattern of symmetry breaking in which we are interested, we must have: (a) at $T=0$ the  stationary point of lowest energy has $\text{SU(2)}_\text{L}\times (1)_Y$  spontaneously broken and (b) $T_H^\ast < T_C^\ast$ so that there is  range of temperature where $\mu_H(T)^2$ is negative and $\mu_C^2(T)$ is positive. For temperatures in this range color is spontaneously broken but $\text{SU(2)}_\text{L}\times U(1)_Y$ is not.

The condition on the temperatures $T_{H,C}^*$ and Eq.~(\ref{eq:pos3}) leads to
\begin{gather}
\nonumber
\frac{ \lambda_H v_H^2}{\left({ \lambda_H} +{\lambda_{HC}}+{3g_2^2/4} +y_t^2 \right)}< \frac{3\mu_C^2}{\left({ \lambda_C} +{\lambda_{HC}/2}+{g_3^2} \right)}\\
  < \frac{6\lambda_{HC} v_H^2}{\left({ 2 \lambda_C} +{\lambda_{HC}}+{2g_3^2} \right)}\,,
\end{gather}
or
\be
\label{eq:break1}
\frac{{ 2 \lambda_C} +{\lambda_{HC}}+{2g_3^2} }{{ \lambda_H} +{\lambda_{HC}}+{3g_2^2/4} +y_t^2 } < \frac{6\lambda_{HC}}{\lambda_H}\,.
\ee
Setting $\lambda_H\simeq 1/4$ as implied by the observed Higgs mass Eq.~(\ref{eq:break1}) then implies 
\be
\label{eq:break2}
\frac{{ 2 \lambda_C} +{\lambda_{HC}}+{2g_3^2} }{1 +{\lambda_{HC}}+{3g_2^2} +4y_t^2 } \lsim {6\lambda_{HC}}\,.
\ee
In short, $\lambda_{HC}$ cannot be arbitrarily small. 

\begin{figure*}
\includegraphics[scale=1.0,width=8cm]{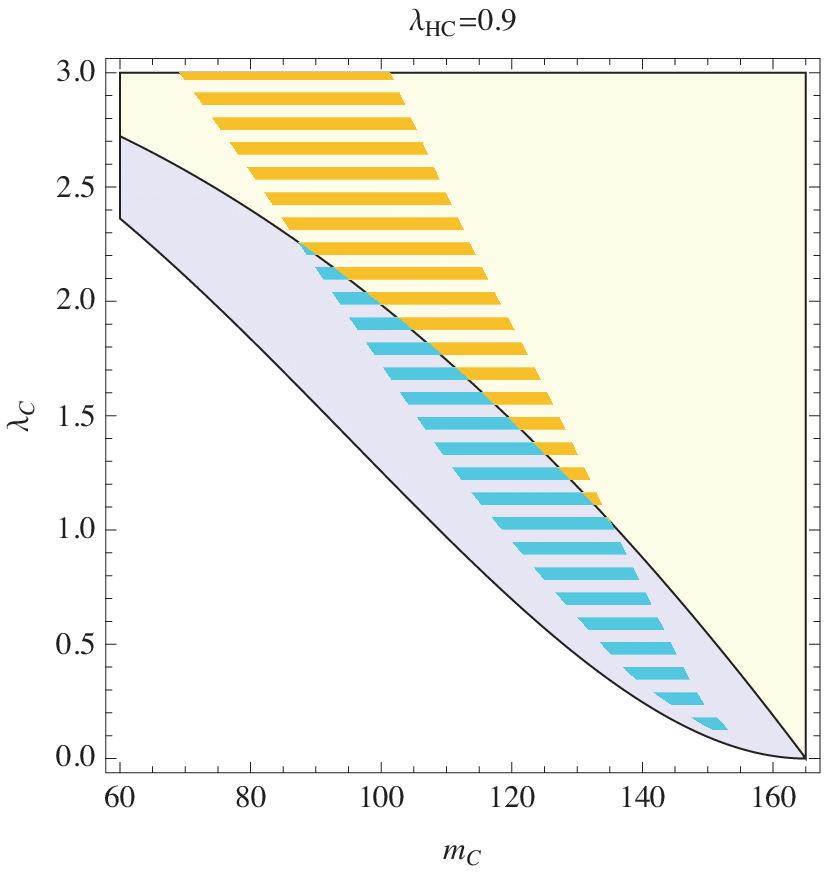}\enspace
\includegraphics[scale=1.0,width=8cm]{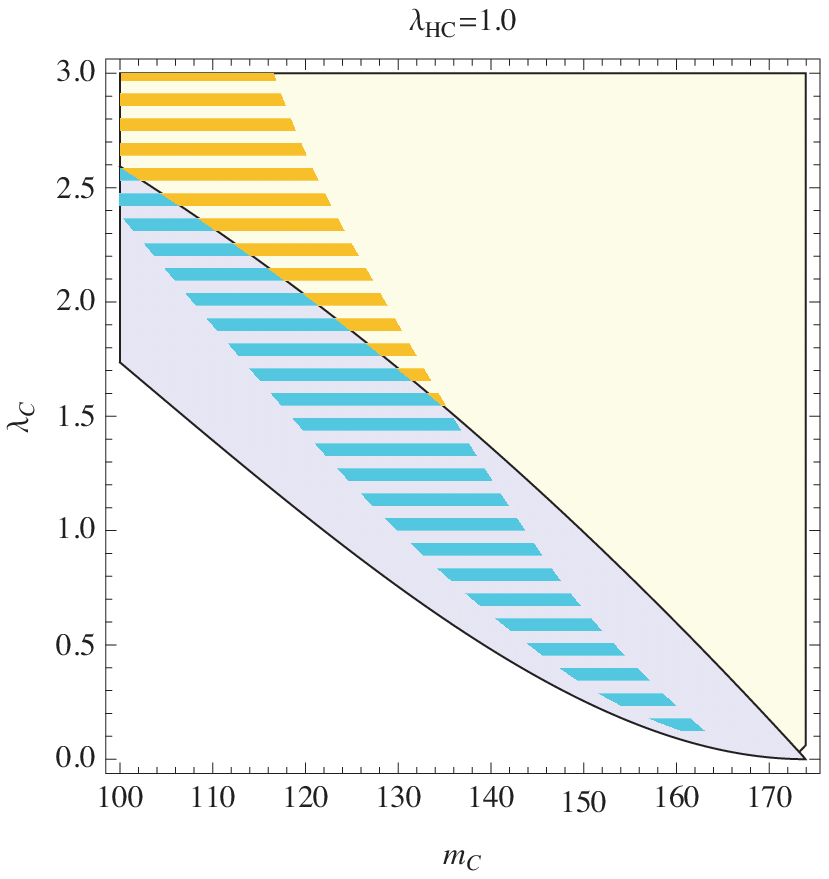}
\caption{Regions in parameter space where color is broken and possibly subsequently restored in the early universe.   Color shading indicates the vacuum structure at zero temperature:  \emph{yellow} representing the region where $\text{SU(3)}_\text{C}$ breaking phase is tachyonic, and \emph{blue} indicating regions where $\text{SU(3)}_\text{C}$ breaking phase is metastable.  In the \emph{white} regions, the standard electroweak phase is metastable, and $\text{SU(3)}_\text{C}$ breaking phase is stable. For $(m_C,\lambda_C)$ inside the  yellow hatched area, the universe first cools into a color-breaking phase, followed by a transition to the color-preserving EW phase. For the blue hatched region, the universe may remain in the color-breaking phase.
{ Left panel} is for $\lambda_{HC}$ = 0.9, and {right panel} is for $\lambda_{HC}$ = 1.0.}
\label{fig:twofield}
\end{figure*}

Now consider the requirement that at zero temperature the energy of the stationary point with only the Higgs getting a vacuum expectation value must be less than the stationary point with only the colored scalar getting an expectation value. This gives,
\begin{eqnarray}
\label{eq:energycond}
\nonumber |E_H| &>& |E_C|\\
\text{or}\qquad{m_H^4 \over 2} &\gtrsim& \frac{(2\lambda_{HC}m_H^2-m_C^2)^2}{2\lambda_C}\,.
\end{eqnarray}
Thus, for a given colored scalar mass $m_C$ and $\lambda_{HC}$, the colored scalar quartic self-coupling $\lambda_C$  also cannot be arbitrarily small. 


To illustrate the range of ($\lambda_C$, $\lambda_{HC}$, and $m_C$) that satisfy the requirements (\ref{eq:break1})-(\ref{eq:energycond}), we plot in Fig.~\ref{fig:twofield} the results of a scan over $\lambda_C$ and $m_C$ for two representative values of $\lambda_{HC}$. In doing so, we have employed the full one-loop effective potential that includes the Coleman-Weinberg terms at $T=0$ and the finite-$T$ contributions without adopting the high-temperature expansion. We have analyzed the evolution of the extrema in a way that maintains gauge-invariance, following the approach outlined in Ref.~\cite{Patel:2011th}. The yellow areas correspond to regions where the EW phase is absolutely stable, and the CoB is a saddle point.  In the blue area, the EW phase is still absolutely stable, but the CoB becomes metastable (no longer a saddle point).  Outside these two regions (white), the EW vacuum is metastable with the CoB phase absolutely stable. For values of the parameters inside the yellow hatched region the universe first cools into a CoB minimum, followed by a transition to the EW, color-preserving vacuum at lower temperatures. On the other hand, inside the blue hatched region it is possible that the universe tunnels from the metastable CoB vacuum to the stable EW phase if the tunneling rate is sufficiently large.

 For purposes of illustration, we focus on the yellow shaded region, wherein color-breaking occurs at moderately high temperatures and not at $T=0$. We find that for fixed $\lambda_{HC}$, we must look to fairly high $\lambda_C$ to find phase transitions involving a CoB phase in the early universe.  As $m_C$ is parametrically increased, the required $\lambda_C$ to obtain a CoB transition is reduced.  
As expected from our general discussion above,  $m_C$ must be relatively light, while $\lambda_C$ is relatively large and may even become non-perturbative.  By increasing $\lambda_{HC}$ the upper limit on $m_C$ increases, as we expect from Eq.~(\ref{bound}). From Eq.~(\ref{eq:energycond}), we also anticipate that as one increases $\lambda_{HC}$, the lower limit on $\lambda_C$ increases as well, a trend we see by comparing the left and right panels of Fig.~\ref{fig:twofield}. In short, it is not possible to make $m_C$ arbitrarily large without entering the realm of non-perturbative quartic couplings.   

It is conceivable that a heavier colored scalar with non-perturbative couplings can lead to the pattern of symmetry-breaking we are interested in, though we cannot say so with confidence based on our perturbative analysis. For $m_C$ on the order of a few times the weak scale, one must confront the present LHC limits on the existence of colored scalars, which are generally precluded for $m_C \lsim 600-900$ GeV, assuming they are pair produced (strongly) and decay semileptonically ({\em e.g.}, first- and second-generation leptoquarks \cite{Aad:2011ch,ATLAS:2012aq,Chatrchyan:2012vza}). 
An exception occurs when the scalar couples strongly only to third generation fermions, as the corresponding mass limits are considerably weaker ({\em e.g.} see Refs.~\cite{Chatrchyan:2012st,Chatrchyan:2012sv} for third generation leptoquarks). We defer the corresponding phenomenological analysis to a future study.


\section{Adding a Singlet}
\label{sec:singlet}
In order to increase the mass of the colored scalar without requiring non-perturbative couplings, we consider augmenting the field content by a real scalar singlet $S$  that obtains a non-zero vev $v_S$ at $T=0$ and generates an additional contribution to $m_C$. Since this vev is not connected with the weak scale, it can be as large as needed to increase $m_C$
without strong coupling. The most general extension of the potential in Eq.~(\ref{eq:modela}) is obtained by adding the potential 
\begin{align}
\label{eq:modelb}
\nonumber\Delta V & =-{\mu_S^2 \over 2} S^2 + {\lambda_S \over 4} S^4 
+\lambda_{HC} (H^{\dagger}H)(C^{\dagger} C)\\
\nonumber&\qquad+\frac{\lambda_{HS}}{2} (H^{\dagger}H)S^2+\frac{\lambda_{CS}}{2}(C^{\dagger} C)S^2\\
&\qquad+{e_S \over 3} S^3 +e_C C^{\dagger} C S + e_H H^{\dagger}H S\,.
\end{align}
Again we take the $\mu^2$'s and the $\lambda$'s positive. The analysis of the vacuum structure its temperature structure is considerably more complicated due to the presence of three rather than two fields.
Before proceeding, however, we make some general observations.

\noindent (1) For a $T=0$ vacuum in which $v_S$ and $v_H$ are both non-zero but color is unbroken, the mass of the colored scalar is given by
\be
\label{eq:mcmodb}
m_C^2 = -\mu_C^2 +\lambda_{HC} v_H^2 + \frac{\lambda_{CS}}{2} v_S^2+e_C v_S\ \ \ .
\ee
In the limit $Z_2:\,S\rightarrow -S$ symmetry (i.e., $e_C=e_S=e_H=0$), it is straightforward to obtain the singlet vev $v_S$:
\begin{equation}
\label{eq:singletvev}
v_S^2={\mu_H^2 \lambda_{HS}-\mu_S^2\lambda_H \over \lambda_{HS}^2/2-\lambda_H\lambda_S}\ \ \ ,
\end{equation}
while the Higgs vev is given by
\begin{equation}
\label{eq:higgsvev}
v_H^2={\mu_S^2 \lambda_{HS}/2-\mu_H^2\lambda_S \over \lambda_{HS}^2/2-\lambda_H\lambda_S}\  \ .
\end{equation}

If in addition to setting the cubic terms to zero we decouple the Higgs doublet from the other fields by setting $\lambda_{HS}=\lambda_{HC}=0$ then the problem of finding the vacuum resembles that of the previous case where the fields were just $H$ and $C$ with the singlet $S$ now playing the role of $H$. Since $v_S$  is not constrained by weak interaction phenomenology it can be  much larger than $v_H$, thereby allowing $m_C$ 
to be greater than a TeV [{\em via} Eq.~(\ref{eq:mcmodb}) ] without the coupling $\lambda_{CS}$ being in the strong coupling regime.

\begin{figure}
\includegraphics[scale=1.0,width=8.3cm]{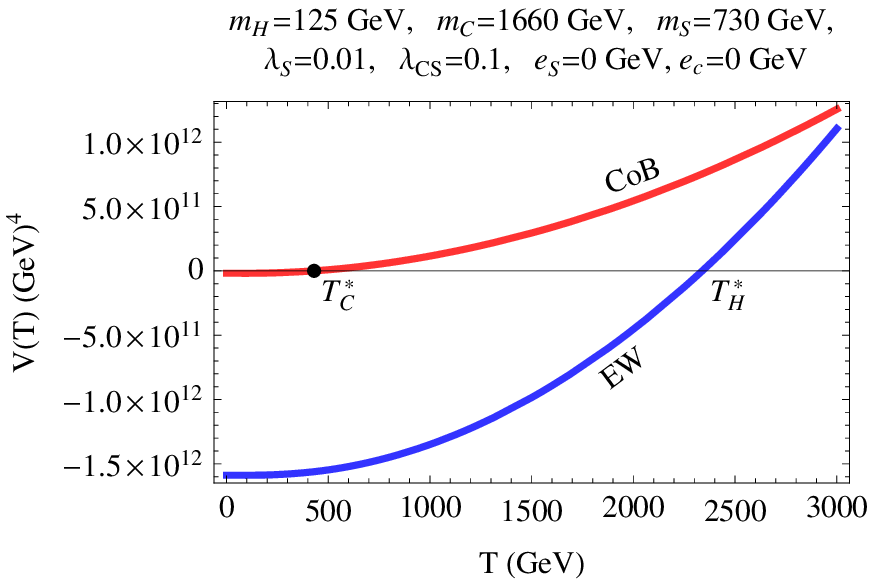}
\caption{The free energy for the color-breaking (CoB) and SM electroweak (EW) phases in the $Z_2$-symmetric limit, as a function of temperature T.  Within this class of parameter choice, the desired pattern of parameter breaking does not happen.}
\label{fig:threefieldZ2Symm}
\end{figure}

\begin{figure}
\includegraphics[scale=1.0,width=8cm]{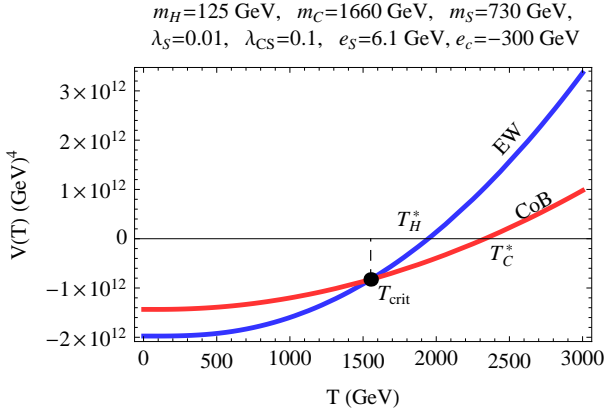}
\caption{The free energy for the color-breaking (CoB) and SM electroweak (EW) phases as a function of T for non-zero $Z_2$ breaking $e$ parameters.  For the choice of parameters labelled above the figure, the critical temperature is $T_{\rm crit} \simeq1550\text{ GeV}$}
\label{fig:threefield}
\end{figure}

\noindent (2) The presence of the additional couplings in Eq.~(\ref{eq:modelb}) will modify the finite-$T$ masses of the scalar fields. Again, for simplicity, we first set the cubic couplings to zero. One then has
\begin{align}
\mu_H^2(T)&=\mu_H^2-T^2\left({ \lambda_H \over 4} +{\lambda_{HC}\over 4}+{\lambda_{HS}\over 24}+{3g_2^2 \over 16} +{y_t^2 \over 4}\right) \\
\mu_C^2(T)& = \mu_C^2-T^2\left({ \lambda_C \over 3} +{\lambda_{HC}\over 6}+{\lambda_{CS}\over 24}+{g_3^2 \over 3} \right) \\
\mu_S^2(T) & = \mu_S^2 -T^2\left({\lambda_S \over 8}+{\lambda_{HS} \over 12}+{\lambda_{CS}\over 8} \right)\,.
\end{align}
From these results, we observe that it is generally difficult to obtain the pattern of symmetry-breaking that we seek. In general, there can exist several vacua at $T=0$, some of which will break $\text{SU(3)}_\text{C}$ but not $\text{SU(2)}_\text{L}$, others breaking $\text{SU(2)}_\text{L}$ and not $\text{SU(3)}_\text{C}$, and still others breaking both ($v_S$ may be non-zero for one or more of these cases). On the one hand, the  color-preserving EW vacua must have the lowest energy at $T=0$. In order for a CoB vacuum to be the state of lowest energy at finite $T$, the energy of the color-preserving vacua must rise more quickly that that of the CoB vacuum. On the other hand, we see that $\mu_{C,H}^2(T)$ will typically decrease faster  as $T$ increases compared to $\mu_S^2(T)$ partly because of its dependence on the gauge couplings. Consequently, as EW symmetry is restored, the color-preserving vacua with $v_s\not=0$ is more likely to emerge as the state of lowest energy than a CoB vacuum.

This situation is illustrated in Fig. \ref{fig:threefieldZ2Symm}, where we show the temperature evolution of the CoB and EW extrema for a representative choice of parameters. While the EW vacuum is the absolute minimum at $T=0$, we observe that one as $T$ increases $\text{SU(3)}_\text{C}$ is restored long before EW symmetry is restored, so that there is never a period when color is broken. 

The inclusion of the cubic terms in the potential (\ref{eq:modelb}) can modify this situation by raising the critical temperature $T_C^\ast$ relative to that for the EW and singlet vacua (For an analogous study of the impact of tree-level cubic terms, see {\em e.g.} Ref.~\cite{Profumo:2007wc}) . This happens for $\lambda_{CS}$ positive as we have chosen it to be. In this case, the energy of the EW minimum  becomes larger than that of the CoB minimum for $T_\mathrm{crit}$. For a range of temperatures satisfying $T_\mathrm{crit} < T < T_C^\ast$, the CoB extremum becomes the state of lowest energy, while for $T>T_C^\ast$, $\text{SU(3)}_\text{C}$ symmetry is restored. 
 
This situation is illustrated in Fig. \ref{fig:threefield}.  Here, we show the  temperature-evolution of the free energy for the phases involved in the transition for two illustrative parameter choices. The   corresponding scalar masses are $(m_C,m_S)\simeq(1660, 730)$ GeV. Since we have set the terms that couple the Higgs doublet to the fields $S$ and $C$ to zero in this example, we can increase the values of $m_C$ and $m_S$ by scaling up the dimensionful parameters $\mu_S^2$ and $\mu_C^2$, $e_S$ and $e_C$ . The evolution of the EW and CoB minima are indicated by the blue and red curves, respectively. In both cases, we have verified that at zero temperature, the color-breaking phase is a saddle point, so that one encounters no possibility that the EW minimum is metastable. 
The pattern of symmetry-breaking can be seen by following the curves from right (high-$T$) to left (low-$T$). At high-$T$, the symmetric phase, corresponding to $V=0$, is the absolute minimum. Below $T_{C}^*$ the CoB minimum takes on negative energy and a transition occurs to the CoB phase. Below the intersection point of the blue and red curves ($T=T_{\rm crit}$), the EW minimum has lower energy and a transition to this phase occurs.

\begin{figure}
\includegraphics[scale=1.0,width=8cm]{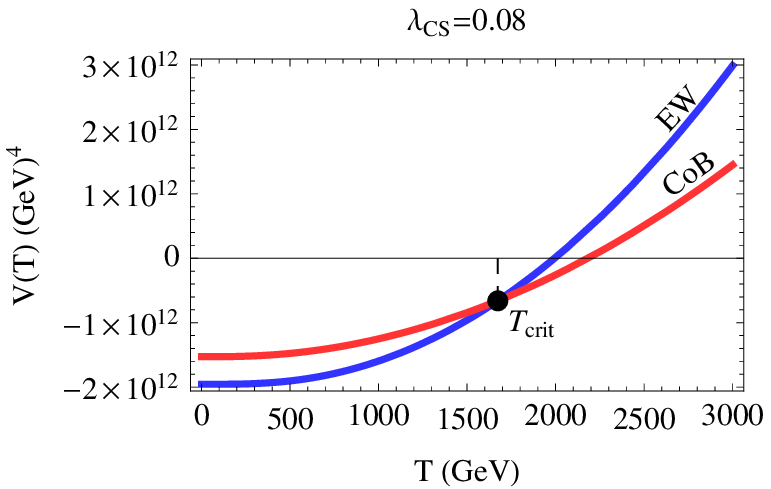}\\
\vspace{5mm}
\includegraphics[scale=1.0,width=8cm]{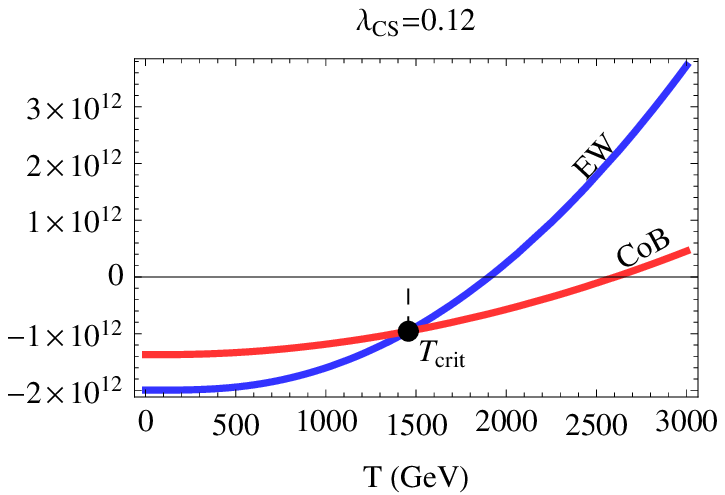}
\caption{The free energy for the CoB and EW phases as a function of T for the same choice of parameters as in Fig. \ref{fig:threefield}, but with coupling constant, \emph{top panel:} $\lambda_{CS}=0.08$ and \emph{bottom panel:} $\lambda_{CS}=0.12$ Raising $\lambda_{CS}$ has the effect of lowering $T_\text{crit}$ and lengthening (in temperature) the duration of spontaneous color-breaking in the early universe.}
\label{fig:threefieldVar}
\end{figure}

A comprehensive study of the parameter space for this multi-field scenario goes beyond the scope of the present work, where we seek to illustrate the basic mechanism. Nonetheless, we have studied the dependence of this pattern on some of the parameters in the potential. Consider changing the value of $\lambda_{CS}$.  The plots in Fig. \ref{fig:threefieldVar} show the impact of increasing or decreasing this coupling by $20\%$. Over this range $T_{\rm crit}$  increases as $\lambda_{CS}$ does. Note also that the color breaking phase exists over a wider range of temperatures as $\lambda_{CS}$ is increased.

Since the masses of the new scalars are much larger than the $m_H$,  one requires fine tuning to maintain a light Higgs. In our case we have implemented this tuning by decoupling the fields.  Moreover, even a modest coupling between the Higgs and the new fields
completely changes the pattern of symmetry breaking. For example if we leave the parameters the same as in Fig. \ref{fig:threefield}, but take $\lambda_{HC}=\lambda_{HS}=0.00012$ the $T=0$ vacuum no longer has electroweak symmetry breaking (i.e. only the singlet has a vev). On the other hand at  $\lambda_{HC}=\lambda_{HS}=0.0001$ the pattern of symmetry breaking is similar to what is shown in Fig. \ref{fig:threefield}. 

We have also studied the dependence of the pattern of symmetry breaking on the cubic couplings $e_C$ and $e_S$ that are essential for the viability of case (b). We find that decreasing the magnitude of $e_S$ (for $e_S>0$) tends to  lower the $T=0$ energy of the CoB extremum, and for sufficiently small $e_S$ the state of minimum energy breaks color. Thus, for a given set of values for the remaining parameters, we expect a lower bound on $e_S$.  Conversely, increasing $|e_C|$ for $e_C<0$ decreases the energy splitting between the $T=0$ EW minimum and the CoB extremum, so we anticipate that in general there will be an upper bound on $|e_C|$ for $e_C<0$ for a given set of the other parameters.

\section{Discussion}
\label{sec:conclude}
We have shown that in extensions of the SM  with an additional color triplet scalar or color triplet scalar and singlet scalar it is possible to have a cosmological history where there is a region of temperature with color spontaneously broken.  At temperatures both above and below this region  region color is restored. In the case of just adding a color triplet to the SM it is not possible (in regions of parameter space where perturbation theory is valid)  for the additional color triplet scalars to be at the TeV scale. Rather, they were necessarily lighter, and  unless one evokes a particular flavor structure in their coupling to fermions such models are ruled out by constraints from the LHC. However by adding another singlet scalar we were able to obtain a region in temperature where color is broken and where the additional colored scalars are sufficiently heavy to avoid LHC constraints. This is the minimal extension of the SM with the novel color breaking cosmological history we are interested in that has the color triplet scalars at the TeV scale (or heavier). In this model we found regions of parameter space with a high temperature color breaking phase but where at zero temperature the color breaking stationary points are not local minima.

Apart from  intrinsic interest in understanding the cosmological evolution of our universe the existence of such a color breaking phase may open up new avenues for generating the baryon excess of the universe. Neglecting Majorana right handed neutrino masses the standard model has a global $B-L$ symmetry. If the new colored scalar is a triplet (under color) and has the right hypercharge to Yukawa couple to a quark-lepton fermion bilinear  then the phase where color is spontaneously broken also breaks the global $B-L$ symmetry.

Finally we would like to mention that adding other degrees of freedom may enlarge the region of parameter space of the scalar potential where the unusual color breaking phase in the early universe occurs. For example if the singlet scalar Yukawa couples to additional fermions then at finite temperature these fermions contribute to the singlet scalar mass squared positively reducing the magnitude of its vacuum expectation value at high temperature compared with the case that the fermions are absent. This could assist the formation of the color breaking phase.

\begin{acknowledgements}
We thank G. Senjanovic for helpful discussions and for pointing out references to some of the earlier literature.
This work of  was supported in part by the U.S. Department of Energy contracts  DE-FG02-08ER41531 (HP and MJRM) and DE-FG02-92ER40701 (MBW) and by the Wisconsin Alumni Research Foundation (HP and MJRM). The research of MBW was supported in part by Perimeter Institute for Theoretical Physics. Research at Perimeter Institute is supported by the Government of Canada and by the Province of Ontario through the Ministry of Economic Development \& Innovation.
\end{acknowledgements}



\bibliographystyle{h-physrev3.bst}
\bibliography{CBrefs.bib}

\begin{thebibliography}{10}

\bibitem{Aad:2012tfa}
ATLAS Collaboration, G.~Aad {\em et~al.},
\newblock Phys.Lett. {\bf B716}, 1 (2012), 1207.7214.

\bibitem{Chatrchyan:2012ufa}
CMS Collaboration, S.~Chatrchyan {\em et~al.},
\newblock Phys.Lett. {\bf B716}, 30 (2012), 1207.7235.

\bibitem{Gurtler:1997hr}
M.~Gurtler, E.-M. Ilgenfritz, and A.~Schiller,
\newblock Phys.Rev. {\bf D56}, 3888 (1997), hep-lat/9704013.

\bibitem{Laine:1998jb}
M.~Laine and K.~Rummukainen,
\newblock Nucl.Phys.Proc.Suppl. {\bf 73}, 180 (1999), hep-lat/9809045.

\bibitem{Csikor:1998eu}
F.~Csikor, Z.~Fodor, and J.~Heitger,
\newblock Phys.Rev.Lett. {\bf 82}, 21 (1999), hep-ph/9809291.

\bibitem{Aoki:1999fi}
Y.~Aoki, F.~Csikor, Z.~Fodor, and A.~Ukawa,
\newblock Phys.Rev. {\bf D60}, 013001 (1999), hep-lat/9901021.

\bibitem{Morrissey:2012db}
D.~E. Morrissey and M.~J. Ramsey-Musolf,
\newblock (2012), 1206.2942.

\bibitem{Land:1992sm}
D.~Land and E.~D. Carlson,
\newblock Phys.Lett. {\bf B292}, 107 (1992), hep-ph/9208227.

\bibitem{Hammerschmitt:1994fn}
A.~Hammerschmitt, J.~Kripfganz, and M.~Schmidt,
\newblock Z.Phys. {\bf C64}, 105 (1994), hep-ph/9404272.

\bibitem{Profumo:2007wc}
S.~Profumo, M.~J. Ramsey-Musolf, and G.~Shaughnessy,
\newblock JHEP {\bf 0708}, 010 (2007), 0705.2425.

\bibitem{Patel:2012pi}
H.~H. Patel and M.~J. Ramsey-Musolf,
\newblock (2012), 1212.5652.

\bibitem{Carena:2008vj}
M.~Carena, G.~Nardini, M.~Quiros, and C.~Wagner,
\newblock Nucl.Phys. {\bf B812}, 243 (2009), 0809.3760.

\bibitem{Stojkovic:2007dw}
D.~Stojkovic, G.~D. Starkman, and R.~Matsuo,
\newblock Phys.Rev. {\bf D77}, 063006 (2008), hep-ph/0703246.

\bibitem{Weinberg:1974hy}
S.~Weinberg,
\newblock Phys.Rev. {\bf D9}, 3357 (1974).

\bibitem{Langacker:1980kd}
P.~Langacker and S.-Y. Pi,
\newblock Phys.Rev.Lett. {\bf 45}, 1 (1980).

\bibitem{Dvali:1995cj}
G.~Dvali, A.~Melfo, and G.~Senjanovic,
\newblock Phys.Rev.Lett. {\bf 75}, 4559 (1995), hep-ph/9507230.

\bibitem{Dvali:1996zr}
G.~Dvali, A.~Melfo, and G.~Senjanovic,
\newblock Phys.Rev. {\bf D54}, 7857 (1996), hep-ph/9601376.

\bibitem{Mohapatra:1979qt}
R.~N. Mohapatra and G.~Senjanovic,
\newblock Phys.Rev.Lett. {\bf 42}, 1651 (1979).

\bibitem{Mohapatra:1979vr}
R.~N. Mohapatra and G.~Senjanovic,
\newblock Phys.Rev. {\bf D20}, 3390 (1979).

\bibitem{Cline:1999wi}
J.~M. Cline, G.~D. Moore, and G.~Servant,
\newblock Phys.Rev. {\bf D60}, 105035 (1999), hep-ph/9902220.

\bibitem{Bimonte:1995xs}
G.~Bimonte and G.~Lozano,
\newblock Nucl.Phys. {\bf B460}, 155 (1996), hep-th/9509060.

\bibitem{Cohen:2011ap}
T.~Cohen and A.~Pierce,
\newblock Phys.Rev. {\bf D85}, 033006 (2012), 1110.0482.

\bibitem{Patel:2011th}
H.~H. Patel and M.~J. Ramsey-Musolf,
\newblock JHEP {\bf 1107}, 029 (2011), 1101.4665.

\bibitem{Aad:2011ch}
ATLAS Collaboration, G.~Aad {\em et~al.},
\newblock Phys.Lett. {\bf B709}, 158 (2012), 1112.4828.

\bibitem{ATLAS:2012aq}
ATLAS Collaboration, G.~Aad {\em et~al.},
\newblock Eur.Phys.J. {\bf C72}, 2151 (2012), 1203.3172.

\bibitem{Chatrchyan:2012vza}
CMS Collaboration, S.~Chatrchyan {\em et~al.},
\newblock Phys.Rev. {\bf D86}, 052013 (2012), 1207.5406.

\bibitem{Chatrchyan:2012st}
CMS Collaboration, S.~Chatrchyan {\em et~al.},
\newblock JHEP {\bf 1212}, 055 (2012), 1210.5627.

\bibitem{Chatrchyan:2012sv}
CMS Collaboration, S.~Chatrchyan {\em et~al.},
\newblock (2012), 1210.5629.

\end{thebibliography}

\end{document}